\documentclass[aps,twocolumn,groupedaddress,amssymb]{revtex4}
\usepackage{latexsym}
\usepackage{amsbsy}
\usepackage{amsmath}
\usepackage{graphicx}
\usepackage{xcolor}
\usepackage{hyperref} % Load hyperref first
\usepackage[capitalise]{cleveref} % Load cleveref last
\hypersetup{
    colorlinks=true,
    linkcolor=black,
    urlcolor=blue,
    citecolor=blue
}
\usepackage{nameref}  % Load nameref after hyperref

\begin{document}

\author{Nazma Firdosh\textsuperscript{1}, Shreyashi Sinha\textsuperscript{1},
Indraneel Sinha\textsuperscript{1}, Mainpal Singh\textsuperscript{2}, Satyabrata Patnaik\textsuperscript{2}, Sujit Manna\textsuperscript{1}\footnote{Contact author: smanna@physics.iitd.ac.in}}
\affiliation{\textsuperscript{1}Department of Physics, {Indian Institute of Technology Delhi}, Hauz Khas, New Delhi 110016, India}
\affiliation{\textsuperscript{2}School of Physical Sciences, Jawaharlal Nehru University, New Delhi, 110067, India}

\preprint{APS/123-QED}

\title{Intertwined magnetic phase driven exchange bias and its impact on the anomalous Hall effect in MnBi$_4$Te$_7$
}

\date{\today}% It is always \today, today,

\begin{abstract}
We report on the interplay between atomic-scale inhomogeneity and competing magnetic phases and its effect on the anomalous Hall effect in the layered antiferromagnet MnBi$_4$Te$_7$, a natural superlattice hosting coexisting ferromagnetic and antiferromagnetic phases. Using a combination of scanning tunneling microscopy (STM), DC and AC magnetization, and magnetotransport measurements, we reveal that intrinsic Mn–Bi antisite defects induce strong interlayer exchange coupling, giving rise to a robust exchange bias observed in both magnetic and Hall responses. The exchange bias undergoes a transition from asymmetric to symmetric behavior between 2 K and 6 K, indicating a temperature-driven dynamical reconfiguration of interfacial spin structures. The training effect analysis revealed a stronger contribution of frozen spins at 2 K compared to 6 K, with relaxation amplitude shift from –264 Oe to +306 Oe. This sign reversal indicates a field-induced change in interfacial coupling. The temperature dependence of longitudinal resistivity (\( \rho_{xx} \)) and magnetization reveals complementary behavior, indicating the coexistence of two distinct spin states near the magnetic transition temperature. The phase-fraction-based resistivity model captures the distinct scattering mechanisms that govern electronic transport across different magnetic regimes. Our findings offer a direct link between microscopic disorder, interfacial magnetism and macroscopic topological phenomena in magnetic topological insulators.

\end{abstract}
\maketitle

%\tableofcontents

\section{Introduction}
The quantum anomalous Hall effect (QAHE) is a striking manifestation of topological quantum physics arising from the interplay of intrinsic magnetic order and strong spin-orbit coupling \cite{Chang2013, Chang2015QAHE,  Serlin2020, Deng2020}. The intrinsic magnetic topological insulator MnBi$_2$Te$_4$(Bi$_2$Te$_3$)$_n$ (where $n = 0, 1, 2, \ldots$) collectively known as the MBT family, has emerged as a rich and tunable platform to explore novel quantum phenomena driven by the interplay of magnetism and topology  \cite{Henk2012, Li2019, He2019TIspintronics, Otrokov2017}. These van der Waals-layered compounds are comprised of alternating magnetic MnBi$_2$Te$_4$ septuple layers(SL) and nonmagnetic Bi$_2$Te$_3$ quintuple layers(QL). They offer a progression of magnetic ground states from antiferromagnetic (AFM) order at $n = 0$ to ferromagnetic (FM) order at $n = 3$, with coexisting AFM and FM phases for intermediate $n = 1$ and $2$ \cite{Klimovskikh2020TunableMagnetism, Tokura2019, Zhao2021}. These magnetic and topological versatility enable access to exotic states such as the quantum anomalous Hall effect (QAHE), axion insulators, Chern insulators, topological magnetoelectric responses, anomalous Nernst effect, topological Hall effect etc. \cite{Deng2020, Ceccardi2023, Li2024Layertronics, ovchinnikov2021intertwined}. This family holds great promise for applications in topological quantum computation, antiferromagnetic spintronics, and 2D spintronics applications. However, many of these remarkable effects have been predominantly observed and characterized in MnBi$_2$Te$_4$ rather than in more structurally homologous members such as MnBi$_4$Te$_7$ \cite{Smejkal2018TopologicalAFM, Baltz2018, Gibertini2019Magnetic2D, Burch2018Magnetism, Smejkal2022AHEAntiferro}.

\begin{figure*}
  \centering
  \includegraphics[width=0.9\linewidth]{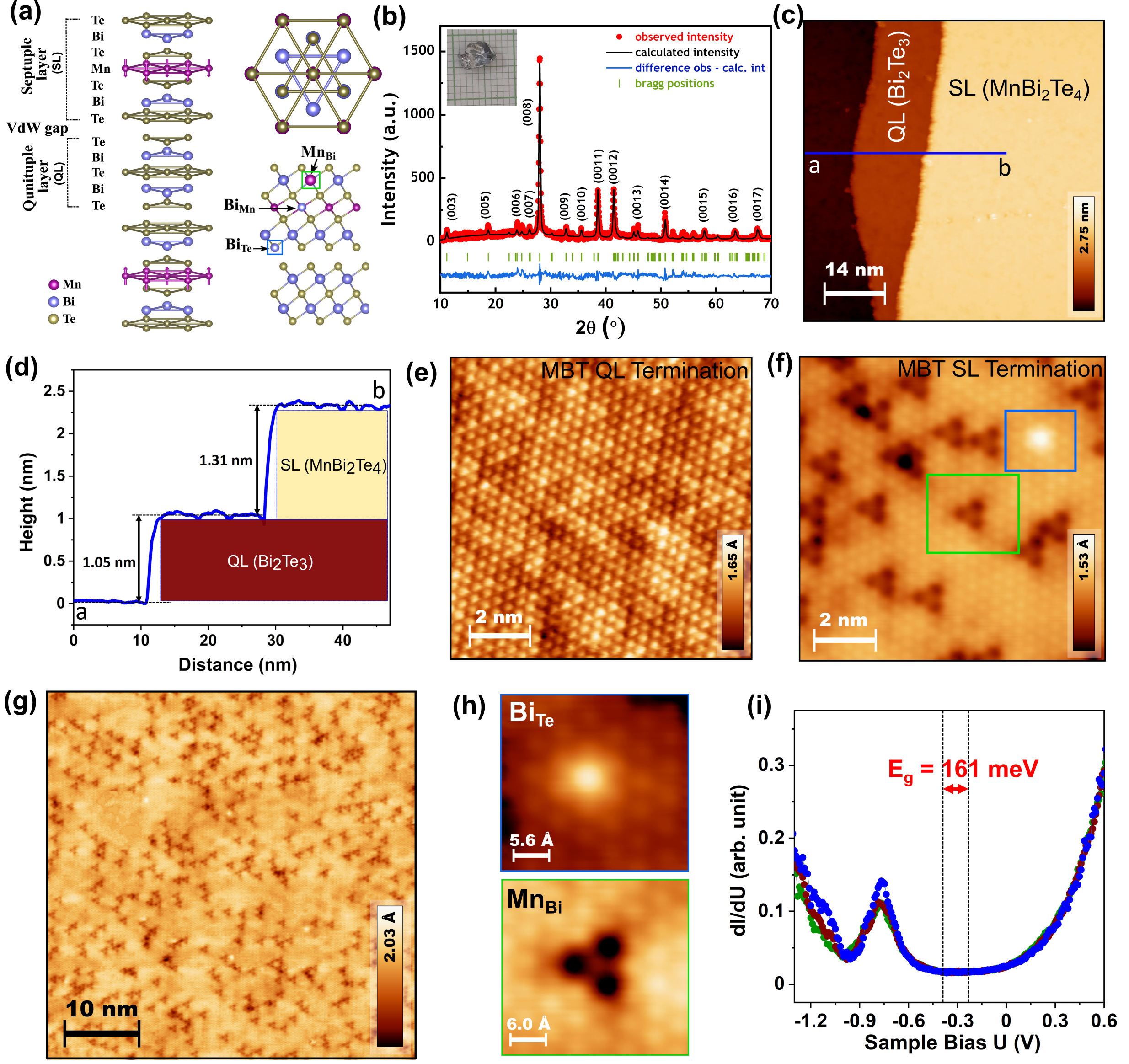}
  \caption{Structural Characterization of MnBi$_4$Te$_7$ single crystal: (a) Crystal structure (side view) of unit-cell consisting of MnBi$_2$Te$_4$ septuple layer(SL) and Bi$_2$Te$_3$ quintuple layers (QL). The pink arrows associated with Mn atomic spin illustrate the interlayer antiferromagnetic (AFM) ordering between layers. Bird view of the crystal structure (right), highlighting the presence of three different defect sites within MnBi$_4$Te$_7$. (b) Rietveld fitted X-ray diffraction pattern confirming the structure of powder form MnBi$_4$Te$_7$ single crystals. (c) Large-scale (70 nm $\times$ 70 nm) STM topography of the Te-terminated (0001) surface of MnBi$_4$Te$_7$ (U = 980 mV, I = 187 pA), displaying two distinct terrace heights corresponding to the quintuple and septuple layers. (d) Line profile across the atomic terraces shows the height variations of each QL and SL layer on the surface. (e) High resolution STM images obtained on the surface of QL area represent the Te-terminated hexagonal surface atomic structure. (f) Atomically-resolved STM image taken (U = 500 mV, I = 476 pA) on septuple layer area shows considerable number of surface and sub-surface defects. (h) Large-scale (50 nm $\times$ 50 nm) STM topography obtained on the septuple layer termination used for estimation of defect concentration. (f) Zoom-in of Fig. 1(f) highlighting the native point defects, including triangular depressions with darker contrast (Mn\textsubscript{Bi}) and circular protrusions exhibiting brighter contrast (Bi\textsubscript{Te}). (i) Differential conductance (dI/dU) vs sample bias revealing a bulk band gap of approximately 161 meV.}
\end{figure*}

As a natural extension of the (MnBi$_2$Te$_4$)/(Bi$_2$Te$_3$) heterostructure series, MnBi$_4$Te$_7$ hosts additional nonmagnetic Bi$_2$Te$_3$ layers. These layers modulate magnetic coupling and potentially reduce the critical field required for spin-flip transitions during magnetic and transport phase studies \cite{Klimovskikh2020TunableMagnetism, Sass2020}. For example, MnBi$_2$Te$_4$ requires fields of several Tesla to achieve ferromagnetic (FM) ordering with an FM transition near 6-8 T and a spin-flop at 3.5 T, imposing significant experimental constraints. In contrast, MnBi$_4$Te$_7$ demonstrates a substantially lower saturation field of 0.22 T at 2 K and a spin-flop transition at 0.04 T \cite{Shi2019}.  This makes MnBi$_4$Te$_7$ a good candidate to explore magnetic and transport phenomena at lower fields. However, MnBi$_4$Te$_7$ suffers from intrinsic limitations such as weak interlayer magnetic coupling and Mn/Bi antisite defects and disorder \cite{garnica2022native, wu2020distinct}. Antisite defects such as Mn$_{Bi}$ and Bi$_{Mn}$ are inherently formed during crystal growth due to the stretching of Mn-Te bonds, increasing the concentration of Mn vacancies \cite{ding2021neutron, wu2020distinct}. Furthermore, the incorporation of large Bi atoms into the structure further increases the strain between Mn layers \cite{Du2021}. These imperfections suppress long-range magnetic order or break essential symmetries required to observe robust quantum states \cite{Yan2022MnBi6Te10}. Thus, it hinders the manifestation and reproducibility of topological effects such as QAHE in bulk crystals, thin films, and flakes \cite{Otrokov2019, Gao2021, Xu2022, shi2024correlation}. In this context, a precise understanding of the defect morphology and its effect on magnetic ground states and topology becomes crucial to bridge the gap between the microscopic (atomic structure and defects) and macroscopic (magnetic and transport) properties of this compound.

The interplay between coexisting magnetic phases in layered magnetic topological materials can be effectively investigated using exchange bias (EB), which captures the interfacial coupling between these competing orders \cite{Paccard1966}. EB manifests as a shift in the magnetic hysteresis loop due to exchange interactions at FM/AFM interfaces, serving as a sensitive tool to probe the underlying magnetic phenomena and interfacial spin dynamics \cite{Xu2022, chong2024intrinsic}. Although EB is extensively studied in engineered FM/AFM heterostructures for spintronics applications, the study of intrinsic exchange bias occurring within a single crystal like MnBi$_4$Te$_7$ remains unexplored \cite{He2018, cui2023layer}. Studying exchange bias in single crystals offers key advantages such as revealing intrinsic spin configurations, avoiding disorder effects introduced by interface fabrication and surface reconstruction. Also, it eliminates the substrate-induced strain and enables depth-resolved insights into magnetic ground states and interfacial spin behavior \cite{Huang2015ExchangeAnisotropy}. Although EB has been reported in thin films of MnBi$_2$Te$_4$, a deeper understanding of related compounds like MnBi$_4$Te$_7$ that host more complex magnetism is still lacking \cite{cui2023layer}. Xu et al. investigated the FM–AFM coexistence in thin exfoliated flakes of MnBi$_4$Te$_7$ using Reflective Magnetic Circular Dichroism (RMCD) measurements \cite{Xu2022}. Despite the potential role of EB in shaping electronic behavior, the direct impact of exchange bias on transport properties in such bulk systems remains insufficiently understood.

In this work, we combine scanning tunneling microscopy (STM) and magneto-transport measurements to investigate the interplay between intrinsic defects, magnetic phase coexistence, and exchange bias in single crystals of MnBi$_4$Te$_7$. Using STM, we investigated the defect morphology and defect density over the top layers of the crystal. Using magnetotransport measurements, we observed a direct and robust correlation between interfacial exchange bias and the shift of anomalous Hall resistance peaks at low temperatures. Moreover, 
training effect analysis further reveals a higher fraction of frozen spin at 2~K compared to 6~K, accompanied by a striking sign reversal in the relaxation amplitude. This suggests a reconfiguration of interfacial spin coupling that dynamically alters the surface electronic states. Also, a crossover from the asymmetric exchange bias at 2~K to a symmetric loop at 6~K reflects a delicate competition between the energy interactions of FM and AFM.
Following the detailed exchange bias study, we also explored the evolution of AFM and FM regions with increasing magnetic field through resistivity--temperature ($\rho$--T) measurements and appropriate fitting models. These results highlight the tunable nature of magnetic interactions in MnBi$_4$Te$_7$ and the interfacial spin dynamics.

\section{Experimental methods}
MnBi$_4$Te$_7$ single crystals were synthesized using the self-flux method \cite{Shi2019}. High-purity Mn, Bi, and Te powders in proper stoichiometry were accurately weighed and ground under an argon atmosphere in a glove box to prevent oxidation. The homogenized mixture was pelletized and sealed in quartz tubes under a vacuum of approximately 10$^{-3}$ Pa. These tubes were then heated to 900°C to ensure complete mixing and homogenization. Following this, the system was cooled rapidly to 600°C, and then slowly to 585°C. The final quenching step in cold water resulted in the formation of plate-like, shiny crystals with dimensions of 7 mm × 6 mm × 2 mm, which exhibited easy exfoliation for subsequent characterization. The structural characterization was conducted using a PANalytical X’Pert X-ray diffractometer(XRD) equipped with Cu K$\alpha$ source ($\lambda$ = 1.5406 Å). Elemental composition and stoichiometry were confirmed by energy-dispersive X-ray (EDX) spectroscopy (Hitachi Table Top TM3000). Scanning tunneling microscopy (STM) measurements were done using a commercial (Unisoku) ultra-high vacuum (UHV) STM operated at liquid nitrogen temperature. To image the MBT crystal surface, the sample was in situ cleaved at base pressure $\sim\,\times\,10^{-10}$~mbar before transfer to the cold STM. All the imaging and spectroscopy (STS) measurements were performed at 78 K using mechanically polished PtIr tips prepared using high-temperature flash heating. STS data was acquired using a lock-in amplifier with 3.6 kHz AC modulation voltage (V\textsubscript{rms} = 20 mV) superimposed on the sample bias.

\begin{figure*}
    \centering
    \includegraphics[width=1\linewidth]{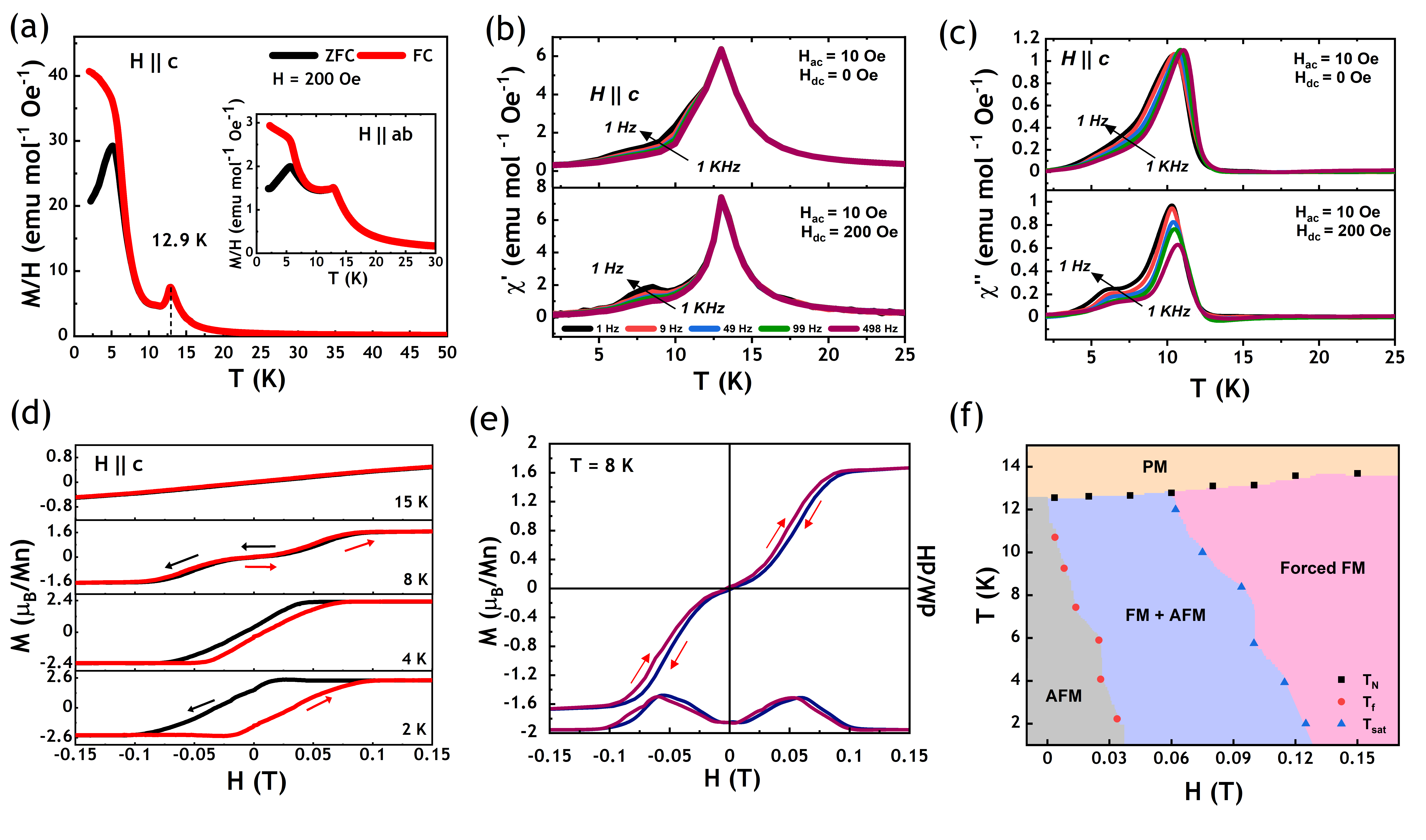}
    \caption{{Temperature and field dependent magnetic properties of MnBi$_4$Te$_7$:} 
    (a)Temperature dependence of magnetization measured under zero-field-cooled (ZFC) and field-cooled (FC) conditions. The sharp peak at 12.9 K corresponds to the antiferromagnetic (AFM) transition, while the low-temperature hump (around 6K) reflects the coexisting FM and AFM magnetic orders. The inset highlights magnetization vs. temperature for $H \parallel ab$; (b) Temperature-dependent AC susceptibility ($H \parallel c$) shows a frequency-independent peak at 12.9 K, confirming stable AFM ordering. A variation between 6–8 K signifies phase coexistence, which intensifies under a DC magnetic field (bottom panel);(d) Isothermal magnetization curves for $H \parallel c$ at different temperatures exhibits spin-flop transitions occurring around 8 K. The zoomed-in image is shown in (e) highights a peak in dM/dH near 0.05 T, indicating a field-induced spin reorientation; (f) Magnetic phase diagram of MnBi$_{4}$Te$_{7}$ for an applied magnetic field H parallel to the c plane. The phase diagram illustrates the antiferromagnetic (AFM), coexisting ferromagnetic (FM) and AFM, forced FM, and paramagnetic (PM) phases as a function of temperature T and magnetic field H. T$_N$ and T$f$ represent the sharp and small peak temperatures in the susceptibility graph at different dc fields respectively while T$_{sat}$ is obtained from the M-H data at various temperatures. }
\end{figure*}

Magnetic measurements were conducted using a Quantum Design Magnetic Property Measurement System (MPMS3) in both zero-field-cooled (ZFC) and field-cooled (FC) modes, with magnetic field swept from -7 Tesla to 7 Tesla over a temperature range of 2K to 300 K. Magnetotransport measurements were performed using a Quantum Design Physical Property Measurement System (PPMS), employing the standard four-probe configuration for electrical characterization.

\section{Results and discussion}
\begin{figure*}
    \centering
    \includegraphics[width=1\linewidth]{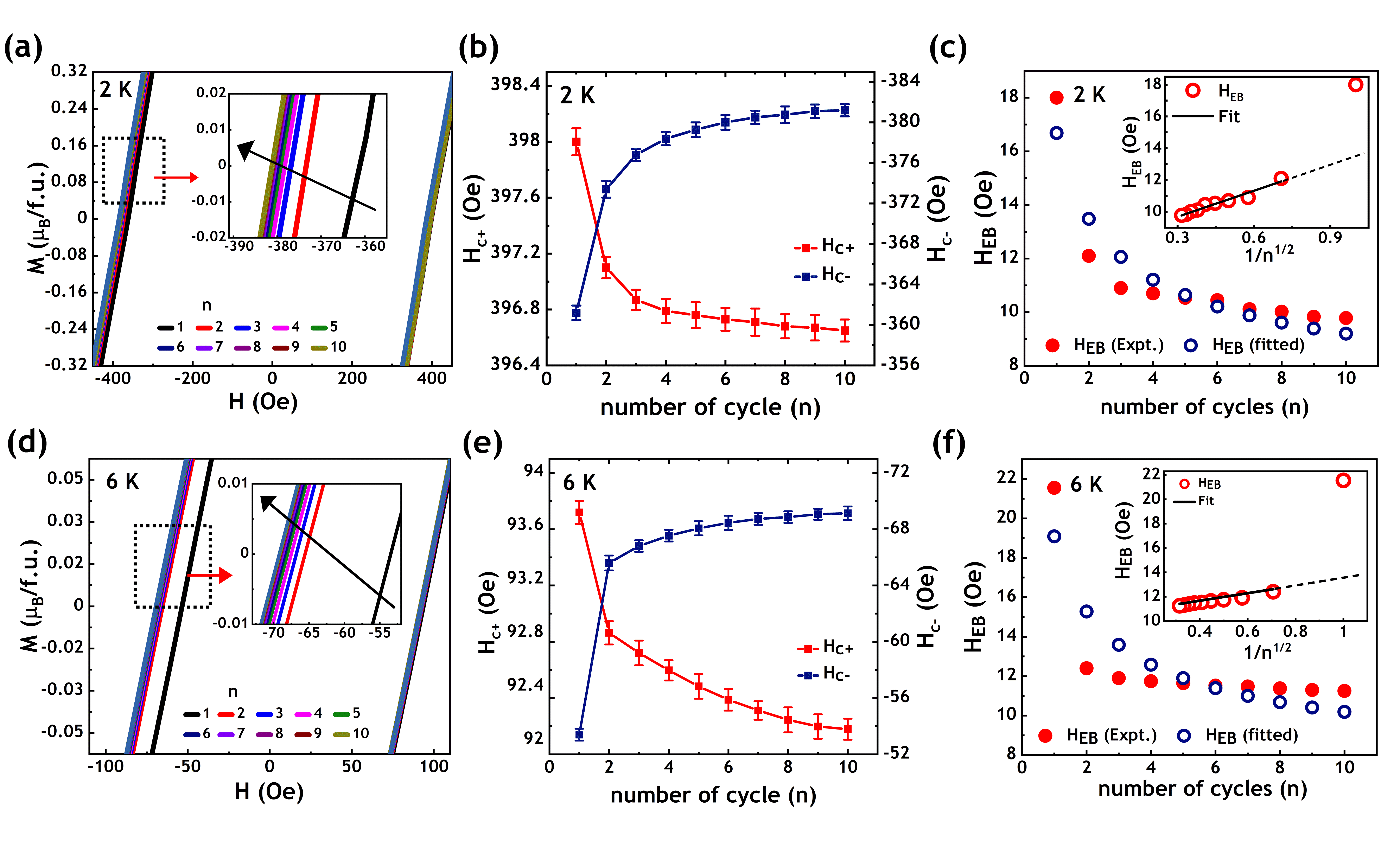}
    \caption{(a) and (d) show isothermal magnetization versus magnetic field (M-H) loops recorded at 2 K and 6 K respectively, after cooling the sample in a 1 T magnetic field over successive field cycles. An enlarged view of the descending branch of the hysteresis loops, showcasing the training effect, is presented in the inset for each respective temperature. (b) and (e) depict the variation in positive and negative coercive fields as a function of the number of field cycles at 2 K and 6 K, respectively. Additionally, (c) and (f) demonstrate the dependence of the exchange bias field (\(H_{\text{EB}}\)) on the number of field cycles, where solid red circles represent experimental data points and empty blue circles denote fitted data points at 2 K and 6 K respectively. The inset figures display the corresponding \(H_{\text{EB}}\) plotted against \(1/n\) and fitted for \(n > 1\). The black solid lines signify the fitted data, while the dashed lines indicate the extrapolation for \(n = 1\).}
\end{figure*}

\subsection{Structural characterization}
MnBi$_4$Te$_7$ crystallizes in the P\(\overline{3}\)m1 space group with alternating unit cells of Bi$_2$Te$_3$ and MnBi$_2$Te$_4$ as shown in Fig. 1(a) \cite{Klimovskikh2020TunableMagnetism}. The septuple-layer (SL) (Te-Bi-Te-Mn-Te-Bi-Te) and the quintuple layer (QL) (Te-Bi-Te-Bi-Te) are held together by van der Waals forces \cite{ding2021neutron, Shi2019}. Magnetism arises from Mn spins, which are arranged in an antiferromagnetic (AFM) interlayer coupling and a ferromagnetic (FM) intralayer coupling, with a perpendicular magneto-crystalline anisotropy (MCA) \cite{Hu2020}. The incorporation of additional Bi$_2$Te$_3$ layers weakens the interlayer AFM coupling compared to MnBi$_2$Te$_4$, leading to a transition from AFM to FM for $n = 3$ in (MnBi$_2$Te$_4$)(Bi$_2$Te$_3$)$_n$ \cite{Klimovskikh2020TunableMagnetism}.  
The XRD pattern obtained from crushed MnBi$_4$Te$_7$ samples is analyzed using the FullProf Rietveld refinement as shown in Fig.~1(b). The sample is found to crystallize in the P\(\overline{3}\)m1 space group, with lattice parameters a = 4.232(4) $\mathrm{\AA}$ and c = 23.247(1) $\mathrm{\AA}$ \cite{Shi2019}. Compositional analysis by energy-dispersive X-ray spectroscopy (EDX) reveals an atomic ratio of 1:3.83:7.06, which is in close agreement with the expected stoichiometric ratio of 1:4:7. Elemental distributions are presented in [Supplementary Fig. S2].  

To further investigate the surface morphology and electronic properties of the material following XRD and EDX measurements, we performed a detailed STM study. Fig. 1(c) and Fig. 1(e–h) present constant-current STM scans of the MnBi$_4$Te$_7$  crystal following in-situ cleaving under ultra-high vacuum conditions. The large-scale STM topography in Fig. 1(c) reveals atomically flat terraces, separated by step heights of approximately 1.31 nm and 1.05 nm, as determined from the line profile shown in Fig. 1(d). The measured height of 1.31 nm corresponds to the septuple layer (SL) height of MnBi$_2$Te$_4$, while the 1.05 nm step height corresponds to the quintuple layer (QL) height of Bi$_2$Te$_3$ \cite{wu2020distinct}. As the crystal cleaves naturally between two adjacent layers, both terraces exhibit Te-terminated surfaces. The hexagonal arrangement of the Te sublattice is clearly visible in the high-resolution STM images of the QL and SL layers (Fig. 1(e) and 1(f), respectively), with a measured in-plane lattice constant of 0.43 nm. Although the surfaces are atomically smooth and clean, various types of native point defects are present. In the SL layer, two predominant defect types are identified (Fig. 1(h)). The first type, which appears as dark triangular features, is attributed to Mn substituting for Bi (Mn\textsubscript{Bi}) in the second layer. The second type, characterized by bright protrusions, is likely associated with Bi replacing top-layer Te atoms (Bi\textsubscript{Te}). Although Mn-substituted defects are frequently observed, Bi-related defects are relatively rare. From the large-scale topography scan (Fig. 1(g)), taken on the SL layer, the coverage area of the defects is calculated. Their coverage areas are approximately 5-7$\%$ and $\leq$1$\%$ respectively. The detailed calculation of defect density \cite{sinha2025insitu} is shown in the Supplementary (Fig. S2). The electronic properties of the MnBi$_4$Te$_7$ surface were further examined through differential conductance (dI/dU) spectroscopy at various locations on the sample. The dI/dU spectra (Fig. 1(i)) exhibit a U-shaped feature with a flat bottom, with a suppression of DOS within the energy range from -214 meV to -375 meV. These energy values correspond to the valence band maximum (VBM) and conduction band minimum (CBM), respectively. The extracted band gap, calculated as the energy difference between these two points, is approximately 161 meV, which is in good agreement with previously reported band gap values \cite{liang2020mapping, wu2020distinct}. This consistency further confirms the high purity of the sample.

\subsection{Isothermal magnetization and the Exchange bias effect}
\subsubsection{Coexisting AFM-FM phase}

To understand the intricate interplay between ferromagnetic and antiferromagnetic interactions in MnBi$_4$Te$_7$, it is essential to probe the thermal evolution of magnetic ordering. This will uncover possible phase coexistence, spin reorientation transitions, and field-sensitive behavior.
To probe this, we performed temperature-dependent magnetization measurements under zero-field-cooled (ZFC) and field-cooled (FC) protocols. Measurements are conducted with the magnetic field applied along both the out-of-plane ($H \parallel c$) and in-plane ($H \parallel ab$) directions, as demostrated in Fig.~2(a).
 The M–T curve shows two pronounced slope changes near 6 K and 13 K, indicating that the system undergoes two distinct magnetic transitions corresponding to different spin configurations. The first transition corresponds to the spin-freezing temperature (\(T_f = 6\,\text{K}\)), below which a FM phase becomes dominant. The second transition at the Néel temperature (\(T_N = 12.9\,\text{K}\)) marks the onset of AFM ordering, in agreement with previous studies~\cite{Shi2019, Klimovskikh2020TunableMagnetism}. In between these two transition temperatures, the Mn spins coexist in mixed FM-AFM states \cite{Ceccardi2023}. To gain deeper insight into the dynamics and the nature of this coexistence, further investigation into the frequency dependence of the magnetic response is required. 
 
 \renewcommand{\arraystretch}{1.5}
\begin{table*}[ht]
\caption{\label{tab: example} Fitted parameters from training effect with two models at 2 K \& 6 K.}
\begin{ruledtabular}
\begin{tabular*}{\textwidth}{@{\extracolsep{\fill}}c p{4cm} p{4cm}} % Adjusted widths for Columns 2 and 3
Model for fitting & T = 2K & T = 6K \\
\hline
Empirical power law\cite{Paccard1966} & H$_{EB}^{\infty}$ : 8.05  \hspace{10pt} $k_H$ 5.45 & H$_{EB}^{\infty}$ : 13.38  \hspace{10pt} $k_H$ 2.77 \\
Frozen and rotatable spin relaxation model\cite{Mishra2009} & H$_{EB}^{\infty}$ : 10.17  \hspace{10pt} $A_f$ -264.26 \hspace{10pt} $P_f$ 0.76  \hspace{20pt} $A_r$ 292.85 \hspace{10pt} $P_r$ 0.76 & H$_{EB}^{\infty}$ : 15.50  \hspace{10pt} $A_f$ 306.53 \hspace{10pt} $P_f$ 0.425 \hspace{20pt} $A_r$ 201.27 \hspace{10pt} $P_r$ 0.42 \\
\end{tabular*}
\end{ruledtabular}
\end{table*}
To confirm the coexistence of FM–AFM ordering and to gain insight into the spin relaxation mechanisms, we measured both the real (\( \chi' (T) \)) and imaginary (\( \chi''(T) \)) parts of the AC susceptibility as functions of frequency and temperature. The real part of AC susceptibility \( \chi'(T) \) shows a prominent frequency-independent peak near 12.9~K, consistent with a conventional antiferromagnetic long-range ordering transition (see Fig.~2(b))~\cite{Balanda2013, Klimovskikh2020TunableMagnetism}.
On the contrary, both $\chi'(T)$ and $\chi''(T)$ exhibit a clear frequency dependence between 6-10 K. This indicates the presence of slow spin dynamics associated with coexisting FM-AFM phases. Below 5~K, the nearly constant $\chi'(T)$ and suppressed $\chi''(T)$ signal suggest the dominance of ferromagnetic ordering or frozen spin dominance. When we apply a finite DC field ($H_\mathrm{dc} = 200$~Oe), the peak position shifts towards the lower frequency, further supporting the field-tunable coexistence of magnetic phases and the presence of spin freezing effects at low temperatures.

Isothermal magnetization $M(H)$ loops measured under both in-plane ($H \parallel ab$) and out-of-plane ($H \parallel c$) magnetic fields are presented in the Supplementary Information (Fig.~S3). M-H loops confirm an out-of-plane easy axis, attributed to the strong interlayer coupling. The zoomed-in M-H loops for ($H\parallel c$), measured in the field range of $\pm0.15$~T, are shown in Fig.~2(d,e). M-H sweeps acquired at 2 K exhibit distinct hysteresis, indicating ferromagnetic behavior of the Mn spins. The ferromagnetic behavior is suppressed, as confirmed by the reduced coercivity observed at 4 K. At 8K, M-H loops show a clear spin flip transition originating from the interlayer AFM coupling.
Ferromagnetic-like hysteresis at low temperatures reflects the anisotropy-driven polarization of magnetic moments in a weakly coupled layered system like MnBi$_4$Te$_7$.  Under \( H\parallel c \), the Mn spins achieve a quick saturation at 0.02 T.

A phase diagram is constructed from AC susceptibility and DC magnetization measurements as shown in Fig.~2(f). The magnetic phase transitions as a function of temperature ($T$) and magnetic field ($H$) identify four distinct regions: antiferromagnetic (AFM), coexisting ferromagnetic and antiferromagnetic (FM + AFM), forced ferromagnetic (Forced FM), and paramagnetic (PM) regions. The transition lines—$T_N$, $T_f$, and $T_{\text{sat}}$—mark the Néel temperature, spin-freezing point, and magnetic saturation, respectively.

The transitions at 12.9 K (AFM ordering) and 6 K (blocking/ spin freezing temperature) enable precise control of magnetic phases through temperature variation, allowing for switching between AFM, FM, or mixed phases. The frequency-dependent susceptibility near 6 K reveals sensitivity to applied frequency, which could be used to fine-tune spin relaxation processes in devices. The interplay between FM and AFM phases may also stabilize specific quantum states, such as topologically protected states, with the help of external fields or exchange bias effects.

\begin{figure*}
    \centering
    \includegraphics[width=1\linewidth]{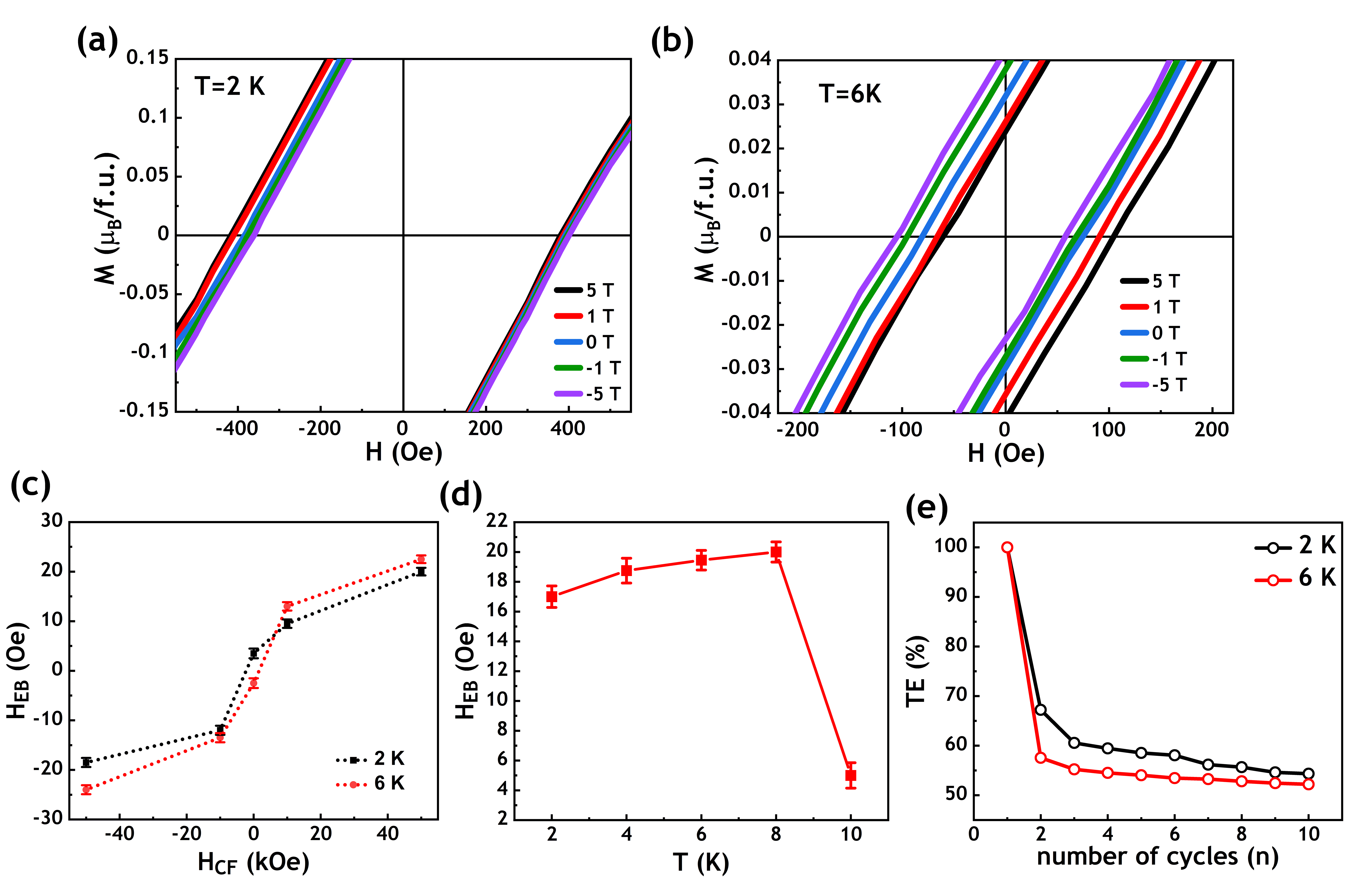}
    \caption {Magnetization versus magnetic field hysteresis loops at (a) 2 K and (b) 6 K, showing the field-induced magnetic response and coercivity characteristics. (c) Exchange bias dependence (\(H_{\text{EB}}\)) on the cooling field (\(H_{\text{CF}}\)) highlighting the relationship between \(H_{\text{CF}}\) magnitude applied during cooling and the resultant exchange bias observed in the magnetic system. (d) Exchange bias field variation with temperature (\(H_{\text{EB}}\)) measured at a constant cooling field of 1 T, indicating an increase in \(H_{\text{EB}}\) up to the blocking temperature (\(T_{\text{B}}\)), followed by a subsequent decline beyond this critical temperature. (e) The percentage reduction in training effect at 2 K and 6 K, quantitatively depicts the influence of temperature on the magnetic training phenomenon, which reflects the stability of the exchange bias over successive field cycling.
    }
\end{figure*}

\subsubsection{Exchange bias and training effect}

We conducted training effect measurements using field-cooled M-H loops with successive cycles to analyze the evolution of exchange bias and interaction parameters \cite{Binek2004}. The M-H loops for 2K and 6K are presented in Fig.~3(a) \& 3(d), respectively. We can see the gradual reduction in the exchange bias effect. The reduction in unidirectional anisotropy with repeated field cycling can be attributed to the demagnetization of AFM spins at the disordered magnetic interface. Coercive fields for the positive and negative branches of M-H loops are shown in Fig.~3(b) \& 3(d) at 2 K and 6 K. We can observe the rapid reduction in the negative coercive field as compared to the positive coercive field. 

The variation of the exchange bias field with the number of loop cycles is presented in Fig.~3(c) \& 3(f) respectively.  Here, the exchange bias field is given by:
$H_{\text{EB}} = ({H_{C+} + H_{C-}})/{2}$ and the coercive field is given by: $H_C={(|H_{C+}| + |H_{C-}|})/{2}$.
To quantitatively comprehend the exchange bias interaction with a successive number of cycles, we fitted the variation of $H_{EB}$ with the number of cycles n with an empirical power law \cite{Paccard1966}.
\begin{equation}
H_n^{EB} - H_{\infty}^{EB} = k_{H}/\sqrt{n} \quad \text{for} \, n > 1,
\end{equation}
Here, H$_{EB}^{\infty}$ denote the values of H$_{EB}$, when n = $\infty$ with constant k$_H$.  The above power law only explains the behavior for $n > 1$ and does not explain the steep decrease between 1\textsuperscript{st} and 2\textsuperscript{nd} loops. The fitted parameters are summarized in Table 1.

We further examined the training effect based on a model that accounts for two distinct relaxation rates associated with frozen and rotatable, uncompensated spin components at the interface. The two spin types are firmly exchange-coupled to the AFM and FM layers within the material respectively, and exhibit distinct relaxation rates. Both are influenced by the reversals of ferromagnetic magnetization, with each contributing uniquely to the training effect \cite{Mishra2009}. Based on this model, the H$_{EB}$ depends on the number of cycles as: 
\begin{equation}
H_n^{EB} = H_{\infty}^{EB} + A_f e^{-n/{p_f}} + A_r e^{-n/{p_r}}
\end{equation}
Here, $A_{f}$ and $A_{r}$ are parameters associated with the contribution to the exchange bias from the "frozen" and "rotational" spins in the system, while $p_{f}$ and $p_{r}$ represent how fast the frozen or rotational spins decay at the FM/AFM interface.

\begin{figure*}[t]
\centering
\includegraphics[width=0.90\linewidth]{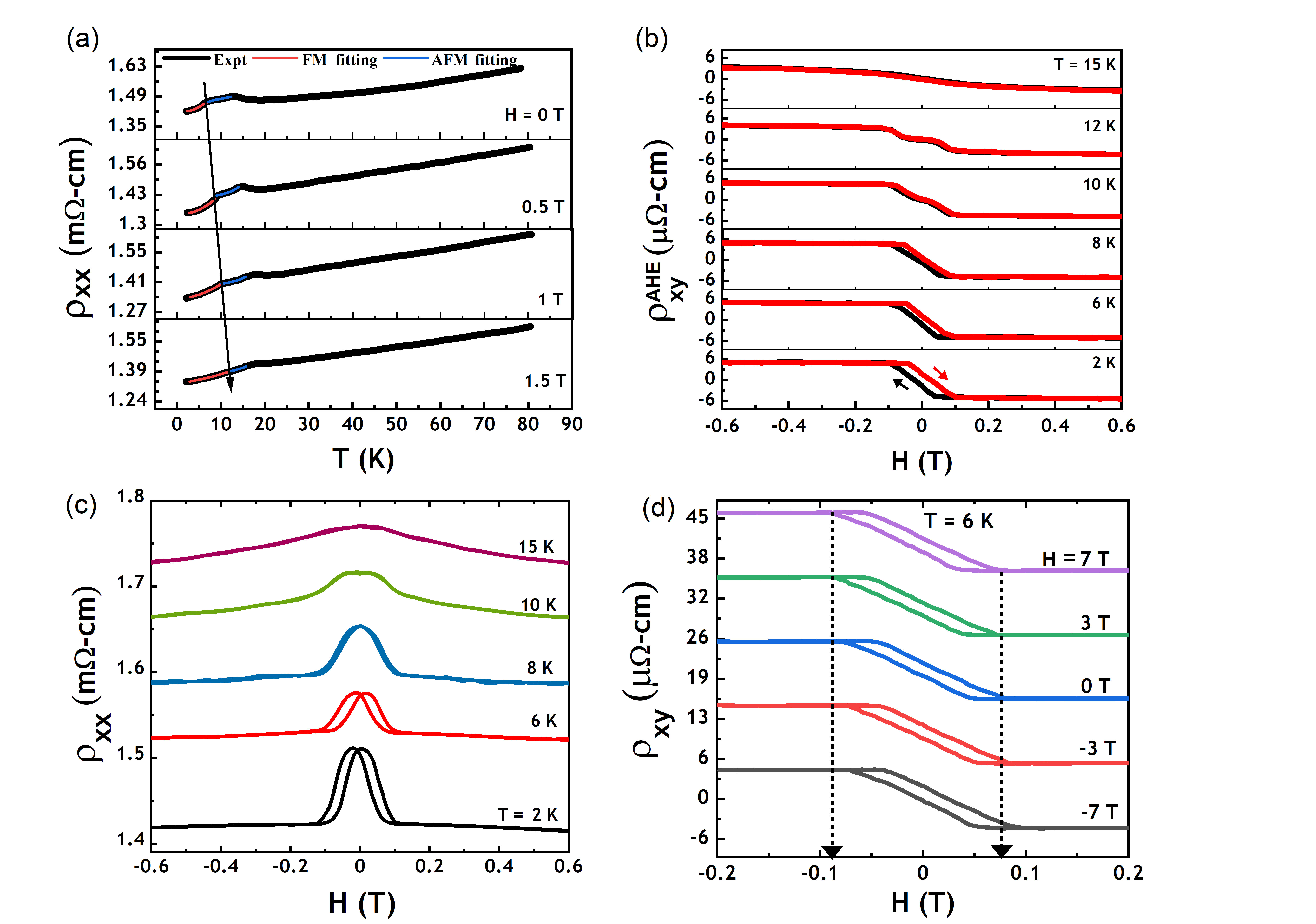}
\caption{(a) Temperature-dependent resistivity of MnBi$_4$Te$_7$ under various magnetic fields. Resistivity fitting in the FM, AFM and PM regions is described by eqn (7), eqn (8) and eqn (9) respectively. These observations indicate an enhancement of ferromagnetic order with an increase in the applied magnetic field. (b) Isothermal Hall resistivity \(\rho_{xy}\) as a function of the magnetic field at various temperatures with \(H\parallel c\). The black and red arrows indicate the direction of the field sweep, and spin-flip transitions are observed in the temperature range of 8–12 K. (c)Isothermal magnetoresistance \(\rho_{xy}\) as a function of the magnetic field at various temperatures with \(H \parallel c\)  (d) Exchange bias observed in the isothermal Hall resistivity at 6 K under different field-cooling protocols, depicts the influence of magnetic history on the Hall response.
}

\end{figure*}
The asymptotic or equilibrium exchange bias field H$_{EB}^{\infty}$ rises from 10.17 at 2 K to 15.50 at 6 K, indicating an increasing exchange bias at 6 K. At 2 K, pinned AFM spins limit coupling with FM spins. As the temperature rises to 6 K, these spins gain thermal energy and become partially unpinned, allowing better alignment with the applied magnetic field and enhancing the exchange interaction. This combination of pinned and unpinned spins at 6 K optimizes exchange bias, with pinned spins providing stability and unpinned spins enhancing coupling, resulting in a stronger exchange bias field as the AFM layer more effectively mediates interactions between the FM layer and the external magnetic field. At 2 K, the system displays ferromagnetic (FM) behavior, with rotatable spins positively contributing to the exchange bias and frozen spins negatively impacting it due to immobility. As the temperature increases to 6 K, a balance shift towards the coexistence of FM and antiferromagnetic (AFM) interactions: frozen spins start to positively influence the exchange bias due to increased thermal energy, while rotatable spins contribute negatively, indicating competition with AFM tendencies \cite{Ali2012}.

Fig.~4(a,b) demonstrates the exchange-bias phenomenon with different field cool cycles at two different temperatures: 2 K (low temperature) and 6 K (near the blocking temperature). At 2 K, an asymmetric shift is observed along the right and left axes, while at 6 K, the shift becomes more symmetric along both axes. This behavior likely reflects the strong pinning effect at low temperatures. The transition from asymmetric to symmetric exchange bias with increasing temperature can be attributed to the gradual unpinning of AFM spins at the FM/AFM interface. Below the blocking temperature, frozen spins dominate, resulting in asymmetry and direction-dependent exchange bias, whereas near the blocking temperature, rotational spins become more prominent, leading to a more balanced and symmetric exchange bias along both axes. Fig.~4(c) shows the variation of exchange bias under different cooling fields for 2K and 6 K. A higher exchange bias is observed at 6K compared to 2 K. Fig.~4(d) shows the variation of the exchange bias field with different temperatures, depicting that H$_{EB}$ increases with the temperature till the blocking temperature, then decreases abruptly at higher temperatures. 

We can see a relative percentage reduction in the training effect with the number of field cycles for the two temperatures in Fig.~4(e). We find that the TE(\%) value decreases to 67\% and 57.5\% of its initial value following the first field cycling (H$_{EB}$) at temperatures of 2 K and 6 K, respectively. Interestingly, although the H$_{EB}$ is lower at 2 K, this temperature exhibits a much more stable EB effect during field cycling when compared to the 6 K sample. For example, after six cycles, the H$_{EB}$ at 2 K reduces to 58\% of its initial value, whereas it retains only 53\% at 6 K. This difference in stability can be attributed to reduced thermal vibrations and enhanced coherence at the lower temperature, which allows for better recovery from cycling perturbations and stronger magnetic interactions within the material.

\subsection{Field-Tuned Transport in Coexisting FM-AFM Phases}
The influence of coexisting phases and the exchange bias effect in MnBi\(_4\)Te\(_7\) should also manifest in the electrical signals. To investigate this, we measured the temperature dependence of longitudinal resistivity ($\rho_{xx}$) under different magnetic fields as shown in Fig.~5a. $\rho_{xx}$  exhibits two distinct transitions at 6 K and 13 K, suggesting the coexistence of FM-AFM magnetic phases. To understand the scattering mechanisms at different temperature regimes, $\rho_{xx}$ has been fitted using the percolation model \cite{Amara2017}. This model accounts for the contributions from ferromagnetic (FM), antiferromagnetic (AFM), and paramagnetic (PM) phases, each characterized by its own temperature-dependent volume fraction.
We define two characteristic crossover temperatures:
    \( T_{\text{[FM-AFM]}} = x \), is the temperature marking the FM to AFM transition, \\
    \( T_{\text{[AFM-PM]}} = y \), is the temperature marking the AFM to PM transition. \\
The evolution of the volume fraction for each magnetic phase is modeled using smooth sigmoid functions. Let \( dx \) and \( dy \) denote the characteristic transition widths near \( x \) and \( y \), respectively. Then, the volume fractions are given by:
{\small
\begin{equation}
    f_{\text{FM}}(T) = \frac{1}{1 + \exp\left(\frac{T - x}{dx}\right)}, \quad
    f_{\text{AFM}}(T) = \frac{1}{1 + \exp\left(\frac{T - y}{dy}\right)}
\end{equation}
\begin{equation}
    f_{\text{PM}}(T) = 1 - f_{\text{FM}}(T) - f_{\text{AFM}}(T)
\end{equation}
}
The total resistivity is then expressed as a weighted sum of the resistivity contributions from each magnetic phase:

\begin{equation}
\resizebox{\linewidth}{!}{$
\rho_{\text{total}}(T) = \rho_{\text{FM}}(T) \cdot f_{\text{FM}}(T)
+ \rho_{\text{AFM}}(T) \cdot f_{\text{AFM}}(T)
+ \rho_{\text{PM}}(T) \cdot f_{\text{PM}}(T)
$}
\end{equation}

Here, \( \rho_{\text{FM}} \), \( \rho_{\text{AFM}} \), and \( \rho_{\text{PM}} \) denote the intrinsic temperature-dependent resistivities associated with the FM, AFM, and PM phases, respectively. These contributions reflect the dominant scattering mechanisms within each magnetic regime. Thus, for our system the total resistivity can be written as:

{\small
\begin{equation}
\begin{split}
\rho_{\text{total}}(T) ={} & \left( \rho_0 + \rho_2 T^2 + \rho_4 T^{4.5} \right) \cdot f_{\text{FM}}(T) \\
&+ \left[ C \cdot T \cdot \exp\left( \frac{D}{k_{\text{B}} T} \right) \right] \cdot f_{\text{AFM}}(T) \\
&+ \left( \rho_{\text{PM0}} + \rho_1 T \right) \cdot f_{\text{PM}}(T)
\end{split}
\end{equation}
}

Based on our electronic transport and magnetic measurements, we identified three distinct regimes: (1) a ferromagnetic (FM) regime between 2-7 K, (2) an antiferromagnetic (AFM) regime between 8-13 K, and (3) a paramagnetic (PM) regime above 13 K. The parameters obtained from fitting the resistivity data with the percolation model are summarized in Table 2.

\begin{table*}[ht]
\caption{ Fitted parameters of temperature-dependent resistivity in different magnetic regions.}
\begin{ruledtabular}
\begin{tabular*}{\textwidth}{@{\extracolsep{\fill}}c c c c c c c c}
\hline
H & C($ 10^{-6} \Omega$-cm) & D/$k_{B}$ (K) & $\rho_{0}$ ($\Omega$-cm) & $\rho_{1}$ ($ 10^{-6} \Omega$-cm $K^{-1}$) & $\rho_{2}$ ($ 10^{-6} \Omega$-cm $K^{-2}$) & $\rho_{PM0}$($\Omega$-cm) & $\rho_{4.5}$ ($ 10^{-11} \Omega$-cm $K^{-2}$) \\
\hline
0 T & 2.128 & 1892 & 0.001375 & 3.4395 & 4.481779 & 0.001475 & -2.35 \\

0.5 T & 2.314 & 1685 & 0.001362 & 3.0128 & 0.213455 & 0.001462 & -1.97 \\
1 T & 2.345 & 1367 & 0.001293 & 2.9265 & 0.113406 & 0.001453 & -1.73  \\
1.5 T & 2.762 & 1198 & 0.001278 & 2.8657 & 0.009436 & 0.001447 & -1.52 \\
\hline
\end{tabular*}
\end{ruledtabular}
\end{table*}

\( \rho_{xx} \) exhibits typical metallic behaviour over the entire temperature range \cite{Shi2019, Klimovskikh2020TunableMagnetism}.
In the low temperature regime (2-7 K), resistivity is strongly influenced by electron-electron scattering. In this region, $\rho_{xx}$ is fitted using the ferromagnetic volume fraction:
\begin{equation}
     \rho_{\text{FM}}=\rho_0 + \rho_2 T^2 + \rho_4 T^{4.5}
\end{equation}
where the coefficients $\rho_0$, $\rho_2$ correspond to residual resistivity, electron–electron (e–e) scattering contribution. $\rho_{4.5}$ represents a combination of electron-electron, electron-magnon (e-m) and electron-phonon (e-p) interactions \cite{hcini2013percolation, Amara2017}.As shown in Fig.~5(a), increasing the magnetic field shifts \(T_{\text{[FM-AFM]}}\) to higher temperatures and simultaneously suppresses \(\rho_{xx}\). This is also evident from the decreasing values of $\rho_2$ and $\rho_{4.5}$ with increasing magnetic field.

Between (8-13 K), the scattering mechanism is dominated by spin fluctuations and is fitted using the following expression
\begin{equation}
 \rho_{_{AFM}} = C \cdot T \cdot \exp\left(\frac{D}{k_{\text{meV}} T}\right)
 \end{equation}
 
where C is a coefficient independent of T and D represents the activation energy \cite{hcini2013percolation}.
 In the antiferromagnetic (AFM) state, spin fluctuations typically become more pronounced as the itinerant carriers interact with local moments. In this region, thermally activated scattering dominates, where charge carriers overcome an activation energy barrier \(D\). The spin fluctuation is suppressed with increasing magnetic field as observed from the fitted value of D for different magnetic fields. Beyond this temperature, \(\rho_{xx}\) shows the typical metallic behaviour and is fitted using the following equation:
\begin{equation}
\rho_{_{PM}} = \rho_{_{PM_0}} + \rho_1 T
\end{equation}
This region is characterized by electron-phonon scattering as the primary mechanism.

The various fitted parameters are mentioned in Table II. The suppression of the intensity of the peak indicates a reduction in the AFM order as the FM phase becomes more dominant. These shifts in peak temperature and intensity with field cooling are characteristic signatures of exchange bias, reflecting changes in the coupling between the FM and AFM phases influenced by the applied field.

The isothermal magnetoresistance (MR) at different temperatures as a function of magnetic fields H $\parallel c$ can be seen in Fig.~5(c). The longitudinal resistivity is symmetrized by averaging over positive and negative magnetic fields to remove the Hall contribution. The nature of the magnetoresistance curve is associated with the AFM-FM spin flop transition due to the strong magnetic field applied along the c-axis \cite{Shi2019, Klimovskikh2020TunableMagnetism}. There appears to be a significant peak around zero magnetic field at temperatures like 8 K and 10 K, which is likely due to enhanced spin scattering in the AFM state. The peak becomes broader and less pronounced at higher temperatures (e.g., 15 K). This suggests that thermal fluctuations destabilize the AFM order.

The Hall resistivity measurements on MnBi$_4$Te$_7$ single crystal with \( H\parallel c \) axis, are shown in Fig.~5(b). The contacts were made in a standard Hall bar geometry. However, small misalignments of the voltage probes can introduce longitudinal resistance components into the transverse resistivity measurement. To minimize this effect, we calculated Hall resistivity \( \rho_{xy} \) using its antisymmetric part:$\rho_{xy}(B) = \frac{1}{2} \left[ \rho_{xy}(+B) - \rho_{xy}(-B) \right]$.\\ This method helps remove contributions that are even in the magnetic field.

For ferromagnetic materials, the transverse resistivity \( \rho_{xy}(B) \) can be written as:
\begin{equation}
\rho_{xy}(B) = \rho_H B + 4\pi \rho_s M
\end{equation}

where \( \rho_H B \) is the ordinary Hall effect due to the Lorentz force, and \( 4\pi \rho_s M \) is the anomalous Hall effect, which depends on the sample's magnetization \( M \). The charge carriers are electrons. The anomalous Hall effect (AHE) can be seen in the magnetic ordering state of MnBi$_4$Te$_7$, the hysteresis loop manifested by the field sweeps during upward and downward directions. Also, the spin-flops are observed at 8K and 10K, which were also reflected in the magnetization versus field plots. The anomalies that occur in the Hall measurements are consistent with the magnetoresistance measurements. The magnetoresistance peaks are observed at the magnetic field, where the Hall resistance switches its value. The anomalous Hall conductivity calculated is 13.6
$\mathrm{\Omega^{-1}cm^{-1}}$. 
The field-dependent anomalous Hall resistivity (\(\rho_{xy}\)) measured at 6 K (Fig.~5(d)) exhibits a clear exchange bias effect, as indicated by the horizontal shift in the hysteresis loops. Notably, \(H_{EB}\) varies with the applied field, with a more pronounced shift under higher fields, indicating field-tunable FM-AFM interactions. This behavior is consistent with magnetization measurements, where exchange bias and training effects further support the presence of pinned and rotatable spin components. The correlation between transport and magnetic data highlights the role of FM-AFM phase coexistence in controlling spin-dependent transport, offering potential insights into stabilizing topological states in quantum anomalous Hall systems.

\section{Conclusion}
In summary, a robust interplay between interfacial magnetism and topological transport in single-crystalline MnBi$_4$Te$_7$ is deduced based on atomic-scale surface imaging, exchange bias and anomalous Hall measurements.  A crossover from asymmetric to symmetric exchange bias, along with temperature-dependent training effects and Hall resistance shifts, reveals a dynamic reconfiguration of interfacial spin coupling. Resistivity measurements under magnetic fields further demonstrate a tunable control over coexisting phases. Our results highlight MnBi$_4$Te$_7$ as a model system where magnetic inhomogeneity and topological properties are intrinsically coupled, offering a pathway to stabilize the quantum anomalous Hall effect via interface engineering—without the need for chemical doping or structural modification. Therefore, our work provides value to be explored for quantum anomalous Hall effect in MBT single crystal which are currently considered only in thin layers. 

\section{ACKNOWLEDGMENTS}
N.F. is grateful for fellowship (DST/INSPIRE/03 /2021/002483) assistance from the Department of Science and Technology, Government of India. This work at IIT Delhi was supported by the ANRF Core Research Grant(CRG/2023/008193). SM, IS and SP acknowledge the partial funding support from the Nano Mission project 'CONCEPT' (DST/NM/TUE/QM-10/2019). The authors acknowledge the Department of physics and the Central Research Facility (CRF), IIT Delhi for providing MPMS, PPMS and STM facilities.

\bibliographystyle{apsrev4-2}
\bibliography{references}

\end{document}